\documentclass[a4paper,12pt,useAMS,numberedappendix]{emulateapj}
\usepackage{apjfonts,graphicx,graphics} 

\usepackage{natbib}
 
\newcommand{\simgt}{\lower.5ex\hbox{$\; \buildrel > \over \sim \;$}}
\newcommand{\simlt}{\lower.5ex\hbox{$\; \buildrel < \over \sim \;$}}
\def\balpha{\mbox{\boldmath $\alpha$}}
\def\bbeta{\mbox{\boldmath $\beta$}}
\def\btheta{\mbox{\boldmath $\theta$}} 
\def\bnabla{\mbox{\boldmath $\nabla$}}
\def\bkappa{\mbox{\boldmath $\kappa$}}
\def\bSigma{\mbox{\boldmath $\Sigma$}}

\def\bd{\mbox{\boldmath $d$}}
\def\bs{\mbox{\boldmath $s$}}

\lefthead{Umetsu et al.}

\righthead{Cluster Mass Profile from the Highest-Quality Lensing Data}  

\begin{document}

\title{A Precise Cluster Mass Profile Averaged from the Highest-Quality
Lensing Data\altaffilmark{*}}

\author{
Keiichi Umetsu\altaffilmark{1}, 
Tom Broadhurst\altaffilmark{2,3},
Adi Zitrin\altaffilmark{4},
Elinor Medezinski\altaffilmark{5},
Dan Coe\altaffilmark{6},
Marc Postman\altaffilmark{6}
} 

\altaffiltext{*}
 {Based in part on data collected at the Subaru Telescope,
  which is operated by the National Astronomical Society of Japan.}
\altaffiltext{1}
 {Institute of Astronomy and Astrophysics, Academia Sinica,
  P.~O. Box 23-141, Taipei 10617, Taiwan.}
\altaffiltext{2}
 {Theoretical physics, University of the Basque Country, Bilbao 48080,
 Spain.}
\altaffiltext{3}
{Ikerbasque, Basque Foundation for Science, Alameda Urquijo, 36-5 Plaza
 Bizkaia 48011, Bilbao, Spain.} 
\altaffiltext{4}
  {School of Physics and Astronomy, Tel Aviv University, Tel Aviv 69978,
 Israel.}
\altaffiltext{5}
{Johns Hopkins University, 3400 North Charles Street, Baltimore, MD
 21218, USA.} 
\altaffiltext{6}
{Space Telescope Science Institute, 3700 San Martin Drive,
Baltimore, MD 21218.}

\begin{abstract}
We outline our methods for obtaining high precision mass
profiles, combining independent weak-lensing distortion, magnification,
and strong-lensing measurements. For massive clusters the strong and
weak lensing regimes contribute equal logarithmic coverage of the radial
profile.  The utility of high-quality data is limited by the cosmic
noise from large scale structure along the line of sight.  This noise is
overcome when stacking clusters, as too are the effects of cluster
asphericity and substructure, permitting a stringent test of theoretical
models.  We derive a mean radial mass profile of four similar mass
clusters of high-quality {\it Hubble Space Telescope} and Subaru images,
in the range $R=40$\,kpc$\,h^{-1}$ to 2800\,kpc\,$h^{-1}$, where the
inner radial boundary is sufficiently large to avoid smoothing from
miscentering effects.  The stacked mass profile is detected at
$58\sigma$ significance over the entire radial range, with the 
contribution from the cosmic noise included. We show that the projected
mass profile has a continuously steepening gradient out to beyond the
virial radius, in remarkably good agreement with the standard
Navarro-Frenk-White form predicted for the family of CDM-dominated halos
in gravitational equilibrium. The central slope is constrained to lie in
the range, $-d\ln\rho/d\ln{r}=0.89^{+0.27}_{-0.39}$.  The mean
concentration is $c_{\rm vir}=7.68^{+0.42}_{-0.40}$ (at $M_{\rm
vir}=1.54^{+0.11}_{-0.10}\times 10^{15}M_\odot\,h^{-1}$), which is high
for relaxed, high-mass clusters, but consistent with $\Lambda$CDM when a
sizable projection bias estimated from $N$-body simulations is
considered.  This possible tension will be more definitively explored
with new cluster surveys, such as CLASH, LoCuSS, Subaru HSC, and XXM-XXL, 
to construct the $c_{\rm vir}$--$M_{\rm vir}$ relation over a
wider mass range.
\end{abstract}
 

\section{Introduction} 
\label{sec:intro}

Clusters of galaxies represent the largest gravitationally-bound objects
in the universe, which contain a wealth of astrophysical and
cosmological information, related to the nature of dark matter,
primordial density perturbations, and the emergence of structure over
cosmic time.  Observational constraints on the properties and evolution
of clusters provide independent and fundamental tests of any viable
cosmology, structure formation scenario, and possible modifications of
the laws of gravity, complementing large-scale cosmic microwave
background and galaxy clustering measurements
\citep[e.g.,][]{Komatsu+2011_WMAP7,Percival+2010_BAO}.

A key ingredient of cluster-based cosmological tests is the mass and
internal mass distribution of clusters.  In this context, the current
cosmological paradigm of structure formation, the standard $\Lambda$
cold (i.e., non relativistic) dark matter ($\Lambda$CDM, hereafter)
model, provides observationally testable predictions
for CDM-dominated halos over a large dynamical range in density and radius.
Unlike galaxies where substantial baryonic cooling is present,
massive clusters are not expected to be
significantly affected by gas cooling
\citep[e.g.,][]{Blumenthal+1986,Broadhurst+Barkana2008}.
This is because the majority of baryons ($\sim 80\%$) in massive
clusters comprise a hot, X-ray emitting diffuse intracluster medium
(hereafter ICM), in which the high temperature and low density prevent
efficient cooling and gas contraction, and hence the gas pressure
roughly traces the gravitational potential produced by the dominant dark
matter \citep[see][]{Kawaharada+2010,Molnar+2010_ApJL}.  The ICM
represents only a minor fraction of the total mass near the centers of
clusters \citep{2008MNRAS.386.1092L,2009ApJ...694.1643U}.  Consequently,
for clusters in a state of quasi equilibrium, the form of their total
mass profiles reflects closely the distribution of dark matter
\citep{Mead+2010_AGN}.

High-resolution $N$-body simulations of collisionless CDM exhibit an
approximately ``universal'' form for the spherically-averaged density
profile of virialized dark matter halos \citep[][NFW
hereafter]{1997ApJ...490..493N}, with some intrinsic variance in the
mass assembly histories of individual halos
\citep{Jing+Suto2000,Tasitsiomi+2004,Navarro+2010}.  The predicted
logarithmic gradient $\gamma_{\rm 3D}(r)\equiv -d\ln{\rho}/d\ln{r}$ of
the NFW form flattens progressively toward the center of mass, with a
central cusp slope flatter than a purely isothermal structure
($\gamma_{\rm 3D}=2$) interior to the inner characteristic radius $r_s
(\simlt 300\,$kpc\,$h^{-1}$ for cluster-sized halos), providing a
distinctive prediction for the empirical form of CDM halos in
gravitational equilibrium.  A useful index of the degree of
concentration is $c_{\rm vir}=r_{\rm vir}/r_s$, which compares the
virial radius $r_{\rm vir}$ to the characteristic radius $r_s$ of the
Navarro-Frenk-White (NFW, hereafter) profile.  This empirical NFW
profile is characterized by the total mass within the virial radius,
$M_{\rm vir}$, and the halo concentration $c_{\rm vir}$.  Theoretical
progress has been made in understanding of the form of this profile in
terms of the dynamical structure using both numerical and analytical
approaches \citep{Taylor+Navarro2001,Lapi+Cavaliere2009,Navarro+2010},
 though we must currently rely on the quality of $N$-body simulations
 when making comparisons with CDM-based predictions for cluster mass
 profiles.

In the context of standard hierarchical clustering models, the halo
concentration should decline with increasing halo mass as dark matter
halos that are more massive collapse later when the mean background
density of the universe is correspondingly lower
\citep{2001MNRAS.321..559B,Zhao+2003,2007MNRAS.381.1450N}.  This
prediction for the halo mass-concentration relation and its evolution
has been established thoroughly with detailed simulations
\cite[e.g.,][]{1997ApJ...490..493N,2001MNRAS.321..559B,2007MNRAS.381.1450N,Duffy+2008,Klypin+2010},
although sizable scatter around the mean relation
is present due partly to variations in formation epoch of halos
\citep{2002ApJ...568...52W,2007MNRAS.381.1450N,Zhao+2009}.  Massive
clusters are of particular interest in this context because they are
predicted to have a relatively shallow mass profile with a pronounced
radial curvature.
Gravitational lensing of background galaxies offers a robust way of
directly obtaining the mass distribution of galaxy clusters
\citep[see][and references
therein]{2001PhR...340..291B,Umetsu2010_Fermi} without requiring any
assumptions on the dynamical and physical state of the system
\citep{Clowe+2006_Bullet,Okabe&Umetsu08}.
A detailed examination of this fundamental prediction has been the focus
of our preceding work
\citep{BTU+05,Medezinski+07,BUM+08,UB2008,2009ApJ...694.1643U,Lemze+2009,Umetsu+2010_CL0024,Umetsu+2011}.

Systematic cluster lensing surveys are in progress to obtain mass
profiles of representative clusters over a wide range of radius by
combining high-quality strong and weak lensing data.
Deep multicolor images of massive cluster cores from Advanced Camera for
Surveys (ACS) observations with the {\it Hubble Space Telescope} ({\it
HST}) allow us to identify many sets of multiple images spanning a wide
range of redshifts for detailed strong-lens modeling
\citep[e.g.,][]{2005ApJ...621...53B,Zitrin+2009_CL0024,Zitrin+2010_A1703,Zitrin+2011_MACS,Zitrin+2010_MS1358,Zitrin+2011_A383}.
The wide-field prime-focus cameras of Subaru and CFHT have been
routinely producing data of sufficient quality to obtain accurate
measurements of the weak-lensing signal, providing model-independent
cluster mass profiles out to beyond the virial radius
\citep[e.g.,][]{BTU+05,BUM+08,2007ApJ...668..643L,UB2008,2009ApJ...694.1643U,Umetsu+2010_CL0024,Umetsu+2011,Coe+2010}.
Our earlier work has demonstrated that without adequate color
information, the weak-lensing signal can be heavily diluted particularly
toward the cluster center by the presence of unlensed cluster members,
leading to biased cluster mass profile measurements with underestimated
concentrations and internal inconsistency, with the weak-lensing based
profile underpredicting the observed Einstein radius
\citep{BTU+05,UB2008,Medezinski+2010}.

Careful lensing work on individual clusters has shown that full mass
profiles constructed from combined strong and weak lensing measurements
show a continuous steepening radial trend consistent with the predicted
form for the family of collisionless CDM halos
\citep{2003A&A...403...11G,BTU+05,BUM+08,UB2008,Umetsu+2010_CL0024,Umetsu+2011}.
Intriguingly these initial results from combined strong and weak lensing
measurements reveal a relatively high degree of halo concentration in
lensing clusters
\citep[e.g.,][]{2003A&A...403...11G,2003ApJ...598..804K,BUM+08,Oguri+2009_Subaru,Zitrin+2011_A383},
lying well above the mass-concentration relation for cluster-sized halos
predicted by the $\Lambda$CDM model, despite careful attempts to correct
for potential projection and selection biases inherent to lensing
\citep{2007ApJ...654..714H,Meneghetti+2010a}.  This apparent
overconcentration of lensing clusters is also indicated by the generally
large Einstein radii determined from strong-lensing data
\citep{Broadhurst+Barkana2008,Meneghetti+2010_MARENOSTRUM,Zitrin+2011_MACS}.

In this paper we explore in greater depth the utility of high-quality
lensing data for obtaining highest-precision cluster mass profiles by
combining all possible lensing information available in the cluster
regime.  This extends our recent weak-lensing work by
\citet{Umetsu+2011}, where a Bayesian method was developed for a direct
reconstruction of the
projected cluster mass profile from  complementary weak-lensing
distortion and magnification effects \citep{UB2008}, the 
combination of which can be used to unambiguously determine the absolute
mass normalization. 
For a massive cluster acting as a super-critical lens, the strong and
weak lensing regimes contribute equal logarithmic coverage of the radial
profile \citep{Umetsu+2011}, so that here we concentrate on those
clusters for which we have high-quality data in both these regimes. 
The high quality of our data is such that we have now become
significantly limited by the cosmic noise from large scale structure
behind the cluster center, where magnified sources lie at greater
distances. This noise is correlated between radial bins, and so can be
overcome by stacking clusters, along independent sight lines.
Stacking also helps average over the effects of cluster
  asphericity and substructure, 
 \citep{Mandelbaum+2006,Johnston+2007_SDSS1,Okabe+2010_WL,Umetsu+2011},
 allowing a tighter comparison of the
  averaged profile with theoretical models.
Our aim here is to develop and apply comprehensive methods to a sample
 of four similarly high-mass lensing
clusters (A1689, A1703, A370, and Cl0024+17), for which we have
previously identified multiply-lensed images  and measured weak
magnification and distortion effects from deep {\it HST} and Subaru
observations
\citep{2005ApJ...621...53B,UB2008,Umetsu+2010_CL0024,Zitrin+2010_A1703,Medezinski+2010,Medezinski+2011,Umetsu+2011}.  

The paper is organized as follows. 
In Section \ref{sec:basis} we briefly summarize the basis of cluster
gravitational lensing.  
In Section \ref{sec:method} we outline our comprehensive methods for
obtaining projected cluster mass profiles from weak-lensing distortion,
magnification, and strong-lensing measurements. 
In Section \ref{sec:app} we apply our methodology to deep {\it
HST} and Subaru observations of four massive clusters to  derive a mean
radial mass profile for the entire clusters, demonstrating how stacking
the weak and strong 
lensing signals improves upon the statistical precision of the mass
profile determination; we then  examine the radial dependence
of the stacked cluster mass profile.
Finally, we discuss our results and conclusions in
\S~\ref{sec:discussion}. 
Throughout this paper, we adopt a concordance $\Lambda$CDM cosmology
with $\Omega_{m}=0.3$,  
$\Omega_{\Lambda}=0.7$, and $h\equiv H_0/(100\, {\rm km\, s^{-1}\,
Mpc^{-1}})=0.7$, unless otherwise noted.

\section{Basis of Cluster Lensing}
\label{sec:basis}

The gravitational deflection of light rays by a cluster 
can be described by the thin lens equation, which relates the angular
position of a lensed image $\btheta$ to the angular position of the
intrinsic source $\bbeta$ as
\begin{equation}
\label{eq:lenseq}
\bbeta=\btheta-\bnabla\psi,
\end{equation}
where $\balpha\equiv \bnabla\psi(\btheta)$ is the deflection field, and
$\psi(\btheta)$ is the effective lensing potential, which is
defined by the two-dimensional Poisson equation as
$\triangle \psi(\btheta)=2\kappa(\btheta)$ with the lensing convergence
$\kappa$ given as a source term.  This equation can be readily inverted
to yield:
$\psi(\btheta)=2\int\!d^2\theta'\,\triangle^{-1}(\btheta,\btheta')\kappa(\btheta')=(1/\pi)\int\!d^2\theta'\,\ln|\btheta-\btheta'|\kappa(\btheta')$, 
so that the deflection field is expressed in terms of $\kappa$ as
\begin{equation}
\label{eq:deflection}
\balpha(\btheta)=\frac{1}{\pi}\int\!d^2\theta'\,\frac{\btheta-\btheta'}{|\btheta-\btheta'|^2} \kappa(\btheta').
\end{equation}
For gravitational lensing in the cluster regime
\citep[e.g.,][]{Umetsu2010_Fermi}, $\kappa$ is expressed as
$\kappa(\btheta)=\Sigma_{\rm crit}^{-1}\Sigma(\btheta)$,
namely the projected mass density $\Sigma(\btheta)$
in units of the critical surface mass density for gravitational
lensing, defined as
\begin{eqnarray} 
\label{eq:sigmacrit}
\Sigma_{\rm crit} = \frac{c^2}{4\pi G D_l} \beta^{-1};
\ \ \ \beta(z_s) \equiv {\rm max}\left[
0,\frac{D_{ls}(z_s)}{D_s(z_s)}\right],
\end{eqnarray}
where $D_s$, $D_l$, and $D_{ls}$ are the proper angular diameter
distances from the observer to the source, from the observer to the
lens, and from the lens to the source, respectively, and 
$\beta=D_{ls}/D_s$ is the angular-diameter distance ratio associated
with the population of background sources.

The deformation of the image for a background source can be described by
the Jacobian matrix
$\cal{A}_{\alpha\beta}\equiv
(\partial\bbeta/\partial\btheta)_{\alpha\beta}=\delta_{\alpha\beta}-\psi_{,\alpha\beta}$  
($\alpha,\beta=1,2$) of the lens mapping,
where $\delta_{\alpha\beta}$ is Kronecker's delta.\footnote{Throughout
the paper we assume in our weak lensing analysis that the angular size of
background galaxy images is sufficiently small 
compared to the scale over which the underlying lensing fields vary, so
that the higher-order weak lensing effects, such as {\it flexion}, can
be safely neglected; see, e.g., \cite{2005ApJ...619..741G,HOLICs1,HOLICs2}.} 
The real, symmetric Jacobian ${\cal A}_{\alpha\beta}$ 
can be decomposed as
${\cal A}_{\alpha\beta} = (1-\kappa)\delta_{\alpha\beta}
 -\Gamma_{\alpha\beta}$,
where $\Gamma_{\alpha\beta}(\btheta) \equiv
(\partial_\alpha\partial_\beta-\delta_{\alpha\beta}\nabla^2/2)\psi(\btheta)$
is the 
trace-free, symmetric shear matrix, 
\begin{eqnarray}
\label{eq:jacob} 
\Gamma_{\alpha\beta}&=&
\left( 
\begin{array}{cc} 
+{\gamma}_1   & {\gamma}_2 \\
 {\gamma}_2  & -{\gamma}_1 
\end{array} 
\right),
\end{eqnarray}
with $\gamma_{\alpha}$ being the components of 
spin-2
complex gravitational
shear $\gamma:=\gamma_1+i\gamma_2$.
In the strict weak-lensing limit where $\kappa,|\gamma|\ll 1$,
$\Gamma_{\alpha\beta}$ induces a quadrupole anisotropy  of the
background image, which can be observed from ellipticities of background
galaxy images.
Given an arbitrary circular loop of radius $\vartheta$ on the sky,
the average tangential shear $\gamma_+(\vartheta)$ around the loop
 satisfies the following identity
\citep[e.g.,][]{Kaiser1995}:
\begin{equation} 
\gamma_+ (\vartheta) = \bar{\kappa}(<\vartheta)-\kappa(\vartheta),
\end{equation}
where $\kappa(\vartheta)$ is the azimuthal average of $\kappa(\btheta)$
around the 
loop, and $\bar{\kappa}(<\vartheta)$ is the average convergence within the
loop. 

The local area distortion due to gravitational lensing, or
magnification, 
is given by the inverse Jacobian determinant,
\begin{equation}
\label{eq:mu}
\mu = 
\frac{1}{|{\rm det}{\cal A}|}
=
\frac{1}{|(1-\kappa)^2-|\gamma|^2|}.
\end{equation}
which can influence the observed surface
density of background sources, expanding the area of sky, and enhancing
the observed flux of background sources \citep{1995ApJ...438...49B}. The
former effect reduces the effective observing  
area in the source plane, decreasing the number of background
sources per solid angle; on the other hand, the latter effect
amplifies the flux of background sources, increasing the number
of sources above the limiting flux. The net effect is known as
magnification bias and depends on the intrinsic slope of the
luminosity function of background sources.

In general, the observable quantity for quadrupole weak lensing
is not the gravitational shear $\gamma$ but the complex {\it reduced}
shear,
\begin{equation}
\label{eq:redshear}
g(\btheta)=\frac{\gamma(\btheta)}{1-\kappa(\btheta)}
\end{equation}
in the subcritical regime where ${\rm det}{\cal A}>0$
(or $1/g^*$ in the negative parity region with ${\rm det}{\cal A}<0$). 
The spin-2 reduced shear $g$ is invariant under the following
global linear transformation:
\begin{equation}
\label{eq:invtrans}
\kappa(\btheta) \to \lambda \kappa(\btheta) + 1-\lambda, \ \ \ 
\gamma(\btheta) \to \lambda \gamma(\btheta)
\end{equation}
with an arbitrary scalar constant $\lambda\ne 0$ 
\citep{1995A&A...294..411S}.
This transformation is equivalent to scaling 
the Jacobian matrix ${\cal A}(\btheta)$ with $\lambda$, 
$\cal {A}(\btheta) \to \lambda {\cal
A}(\btheta)$, and hence leaves the critical curves ${\rm det}{\cal
A}(\btheta)=0$ invariant.  
Furthermore, the curve $\kappa(\btheta)=1$, on which the gravitational
distortions disappear, is left invariant under the transformation
(\ref{eq:invtrans}).

This mass-sheet degeneracy can be unambiguously broken
by measuring the magnification effects,
because the magnification $\mu$ transforms under the invariance
transformation 
(\ref{eq:invtrans}) as 
\begin{equation}
\mu(\btheta) \to \lambda^2 \mu(\btheta).
\end{equation}
In practice, the lens magnification $\mu$ can be measured from
characteristic variations in the number density of background galaxies
due to magnification bias~\citep{1995ApJ...438...49B,Umetsu+2011} as
\begin{equation}
n_\mu(\btheta)=n_0\mu(\btheta)^{2.5s-1},
\end{equation}  
where $n_0=dN_0(<m_{\rm cut})/d\Omega$ is the unlensed number density of
background sources for a 
given magnitude cutoff $m_{\rm cut}$, approximated locally as a
power-law cut with slope $s=d\log_{10} N_0(<m)/dm$ ($s>0$).
In the strict 
weak-lensing limit, the magnification bias is $\delta n_\mu/n_0\approx
(5s-2)\kappa$. 
For red background galaxies the intrinsic count slope $s$ at faint
magnitudes is relatively flat, $s\sim 0.1$,
so that a net count depletion results
\citep{BTU+05,UB2008,Umetsu+2010_CL0024,Umetsu+2011}.  On
the other hand, the faint blue background population tends to have a
steeper intrinsic count slope close to the lensing invariant slope
($s=0.4$).
Alternatively, the constant $\lambda$ can be determined such that
the mean $\kappa$ averaged over the outermost cluster region
vanishes, if a sufficiently wide sky coverage is available.\footnote{Or,
one may constrain the constant $\lambda$ such that the enclosed mass
within a certain aperture is consistent with cluster mass
estimates from some other observations
\citep[e.g.,][]{Umetsu+Futamase1999}.}

\section{Cluster Lensing Methodology}
\label{sec:method}

In this section we outline our methods for obtaining cluster mass
profiles in a continuous radial coverage from the central region to
beyond the virial radius, by combining independent
weak-lensing  distortion, magnification, and strong-lensing
measurements. 

\subsection{Cluster Weak Lensing}
\label{subsec:wl}

The relation between observable distortion ($g$) and underlying
convergence ($\kappa$) is
nonlocal, and the convergence derived from distortion data alone
suffers from a mass-sheet degeneracy (\S~\ref{sec:basis}). However, by
combining the complementary distortion ($g$) and magnification ($\mu$)
measurements 
the convergence can be obtained unambiguously with the correct
mass normalization.

We construct a discrete convergence profile in the weak-lensing
regime from observable lens distortion and magnification profiles,
$g_+(\theta)=\gamma_+(\theta)/[1-\kappa(\theta)]$ 
and $n_\mu (\theta) = n_0\mu(\theta)^{2.5s-1}$
\citep[see Section 3 and Appendix B of][for details of weak-lensing profile
measurements]{Umetsu+2011}, following the 
Bayesian prescription given by \citet{Umetsu+2011}. The Bayesian
approach allows  
for a full parameter-space extraction of model and calibration parameters.
A proper Bayesian statistical analysis is of particular importance to
explore the entire parameter space and investigate the parameter
degeneracies, arising in part from the mass-sheet degeneracy.  

In the Bayesian framework, we sample from the posterior probability
density function (PDF) of the underlying signal  $\bs$ given the data
$\bd$, $P(\bs|\bd)$.  
Expectation values of any statistic of the signal $\bs$ shall converge
to the expectation values of the a posteriori marginalized PDF,
$P(\bs|\bd)$.  The variance covariance matrix $C$ of $\bs$ is
obtained from the resulting posterior sample.
In our problem,
the signal $\bs$ is a vector containing 
the discrete convergence profile,  
$\kappa_i\equiv \kappa(\theta_i)$ with $i=1,2,..,N^{\rm wl}$ in the
weak-lensing regime ($\theta_i>\theta_{\rm Ein}$), and the average
convergence within 
the inner radial boundary $\theta_{\rm min}^{\rm wl}$ of the weak
lensing data,  $\overline{\kappa}_{\rm
min}\equiv \overline{\kappa}(<\theta_{\rm min}^{\rm wl})$, so that $\bs
=\{\overline{\kappa}_{\rm min},\kappa_i\}_{i=1}^{N^{\rm wl}}$, being 
specified by $(N^{\rm wl}+1)$ parameters. 
The Bayes' theorem states that
\begin{equation}
P(\bs|\bd) \propto P(\bs) P(\bd|\bs),
\end{equation}
where ${\cal L}(\bs)\equiv P(\bd|\bs)$ is the 
likelihood of the data
given the model ($\bs$), and $P(\bs)$ is the prior probability
distribution for the model parameters.
The ${\cal L}(\bs)$ function for
combined weak lensing observations is given as a product of the
two separate likelihoods, ${\cal L}_{\rm wl}={\cal L}_g{\cal L}_\mu$,
where ${\cal L}_g$ and ${\cal L}_\mu$ are the likelihood functions for
distortion and magnification, respectively, as given in
\citet{Umetsu+2011}.   
The log-likelihood for combined weak-lensing distortion and
magnification observations, $\{g_{+,i}\}_{i=1}^{N^{\rm wl}}$ and
$\{n_{\mu,i}\}_{i=1}^{N^{\rm wl}}$, is given as
\begin{equation}
-2\ln{\cal L_{\rm wl}}=\displaystyle\sum_{i=1}^{N^{\rm wl}}
\frac{[g_{+,i}-\hat{g}_{+,i}(\bs)]^2}{\sigma_{+,i}^2}+
\displaystyle\sum_{i=1}^{N^{\rm wl}}
\frac{[n_{\mu,i}-\hat{n}_{\mu,i}(\bs)]^2}{\sigma_{\mu,i}^2},
\end{equation}
where $(\hat{g}_{+,i},\hat{n}_{\mu,i})$ are the theoretical
predictions for the corresponding observations; the errors
$\sigma_{+,i}$ for $g_{+,i}$ $(i=1,2,...,N^{\rm
wl})$ due primarily to the variance of the intrinsic source
ellipticity distribution can be conservatively estimated from the data
using bootstrap techniques; the errors $\sigma_{\mu,i}$ for $n_{\mu,i}$
($i=1,2,...,N^{\rm wl}$)
include both contributions from Poisson errors in the counts
and contamination due to intrinsic clustering of red background
galaxies \citep{Umetsu+2011}.

For each parameter of the model $\bs$, we consider a simple flat prior with a
lower bound of $\bs=0$, that is,
$\overline{\kappa}_{\rm min}>0$ and $\kappa_i >0$.
Additionally, we account for the calibration uncertainty in the
observational parameters, such as 
the normalization and slope parameters ($n_0,s$) of
the background counts and the relative lensing depth 
due to population-to-population variations between the background
samples used for the magnification and distortion measurements
\citep[see][]{Umetsu+2011}.

\subsection{Cluster Strong Lensing}
\label{subsec:sl}

We apply our well-tested approach to strong-lens modeling,
which has previously uncovered large numbers of
multiply-lensed galaxies in ACS images of many clusters, such as
A1689 at $z=0.183$
\citep{2005ApJ...621...53B}, 
Cl0024+17 at $z=0.395$ \citep{Zitrin+2009_CL0024}, 
12 high-$z$ MACS clusters \citep{Zitrin+2011_MACS}, 
MS 1358+62 at $z=0.33$ \citep{Zitrin+2010_MS1358},
and A383 at $z=0.188$ \citep{Zitrin+2011_A383}.
Briefly, the basic assumption adopted is that mass
approximately traces light, so that the photometry of the
red cluster member galaxies is used as the starting point
for our model. Cluster member galaxies are identified
as lying close to the cluster sequence by {\it HST} multiband photometry.

In the strong-lensing regime 
we approximate the large scale distribution of cluster
mass by assigning a power-law mass profile to each
cluster galaxy, the sum of which is then smoothed.
The degree of smoothing ($S$) and the index
of the power-law ($q$) are the most fundamental parameters determining
the cluster mass profile dominated by dark matter.
A worthwhile improvement in fitting the location of the lensed images
is generally found by expanding to first order the gravitational
potential of this smooth component, equivalent
to a coherent external shear $\Gamma^{\rm ex}_{\alpha\beta}$
($\alpha,\beta=1,2$) 
describing the overall matter ellipticity.
The direction $\phi_{\rm ex}$ of the spin-2 external shear $\Gamma^{\rm
ex}_{\alpha\beta}$ and its
amplitude $|\gamma_{\rm ex}|$ are free parameters, allowing for some
flexibility in the relation 
between the distribution of dark matter and the distribution of
galaxies, which cannot be expected to trace each other
in detail.

The total deflection field 
$\balpha(\btheta)=\sum_j\balpha_j(\btheta)=(\Sigma_{\rm crit}^{-1}/\pi)\int\!d^2\theta'\,(\btheta-\btheta')/|\btheta-\btheta'|^2\sum_j\Sigma_j(\btheta')$  
consists of the galaxy 
component $\balpha_{\rm gal}(\btheta)$, scaled by a factor $K$, the
smooth cluster dark-matter component 
$\balpha_{\rm DM}(\btheta)$, scaled by $(1-K)$, and the external-shear
component $\balpha_{\rm ex}(\btheta)$,
\begin{equation}
\balpha(\btheta)=K\balpha_{\rm gal}(\btheta)+(1-K)\balpha_{\rm
 DM}(\btheta)
+\balpha_{\rm ex}(\btheta),
\end{equation}
where 
$\alpha_{{\rm ex},\alpha}(\btheta)=\left(\Gamma^{\rm
ex}\right)_{\alpha\beta}\Delta\theta_\beta$ 
with $\Delta\btheta$ being the displacement vector of the
angular position $\btheta$ with respect to a fiducial reference
position. 
The overall normalization ${\cal N}$ of the model and the relative
scaling $K$ of the smooth dark matter component versus the galaxy
contribution bring the total number of free parameters in
the model to 6. This approach to strong lensing is sufficient to accurately
predict the locations and internal structure of
multiple images, since in practice the number of multiple
images uncovered readily exceeds the number of free
parameters, so that the fit is fully constrained.

We use this preliminary model to delens the more obvious
lensed galaxies back to the source plane by subtracting
the derived deflection field. We then relens the source
plane in order to predict the detailed appearance and location
of additional counter images, which may then be
identified in the data by morphology, internal structure
and color. The best-fit strong-lensing model is assessed by minimizing
the $\chi^2$ value in the image plane:
\begin{equation}
\chi^2_{\rm sl}=\displaystyle\sum_i{
\frac{[\btheta_i-\hat{\btheta}_i(q,S,{\cal N},K,\Gamma^{\rm
ex})]^2}{\sigma_i^2}},
\end{equation}
where $i$ runs over all lensed images,
$\hat{\btheta}_i(q,S,{\cal N},\Gamma^{\rm ex})$ is the position
given by the model, $\btheta_i$ is the observed image 
position, and $\sigma_i$ is the positional measurement error.
For each model parameter, we estimate the $1\sigma$ uncertainty by
$\Delta\chi^2\equiv \chi^2-\chi^2_{\rm min}=1$ in the six-parameter
space. 
The uncertainties for the $\Sigma(\btheta)$
field and the $\Sigma(\theta)$ profile are estimated by propagating the
errors on the strong-lens model parameters, $(q,S,{\cal
N},K,\Gamma^{\rm ex})$.

\subsection{Combining Weak and Strong Lensing}
\label{subsec:wl+sl}

We derive a full-radial mass profile on an individual cluster basis by
combining independent weak and strong lensing data, which can be
compared for consistency in the region of overlap.
In order to obtain meaningful radial profiles, one
must carefully define the center of the cluster. It is often assumed
that the cluster mass centroid coincides with the position
of the brightest cluster galaxy (BCG), whereas the BCGs can be
offset from the mass centroids of the corresponding dark matter
halos \citep{Johnston+2007_SDSS1,Oguri+2010_LoCuSS,Oguri+Takada2011}.
\citet{Umetsu+2011} adopted the location of the BCG 
as the cluster center in their one-dimensional profile analysis of five
massive clusters.
A small offset of typically $\simlt 20$\,kpc\,$h^{-1} \equiv d_{\rm
off}$ is found by \citet{Umetsu+2011} between the BCG and the
dark matter center of mass recovered from strong-lens modeling (Section
\ref{subsec:sl}).
In the following we will adopt the BCG position as the cluster center, and    
limit our analysis to radii greater than $R_{\rm
min}\equiv 2d_{\rm off}=40\,$kpc\,$h^{-1}$, beyond which the
cluster miscentering effects on the $\Sigma$ profile are negligible
\citep[see Section 10 of][]{Johnston+2007_SDSS1}.

Having defined the cluster center, we can construct a joint discrete
mass profile $\bSigma =\{\Sigma(R_i)\}_{i=1}^{N}$ as a function of the
projected radius $R=D_l\theta$ by combining the weak and strong lensing
$\kappa$ profiles:  
$\Sigma(R_i)= w_i^{-1}\kappa(\theta_i)$ ($i=1,2,...,N$), where $w_i$ is the
lensing efficiency function, or the inverse critical surface mass density, 
$w_i \equiv (\Sigma_{{\rm crit},i})^{-1} = (4\pi G/c^2) D_l \beta_i$, 
Note, the $i$ dependence arises 
because strong and weak lensing profiles with different depths are
combined together.  
To simplify the analysis, we exclude 
the strong-lensing data points in the region of overlap
(typically, $\theta_{\rm Ein}\simlt \theta\simlt 2\theta_{\rm Ein}$)
as well as  the central weak-lensing bin
$\overline{\kappa}_{\rm  min}$, when defining the joint $\Sigma$ profile.

The formulation thus far allows us to derive
covariance matrices $C^{\rm stat}_{ij}$ of
statistical measurement errors for individual cluster $\bkappa$
profiles. Here we take into account the effect of uncorrelated large
scale structure projected along the line of sight on the 
error covariance matrix $C^{\rm lss}_{ij}$ as $C=C^{\rm stat}+C^{\rm
lss}$, where $C^{\rm lss}$ is given as \citep{1998MNRAS.296..873S,2003MNRAS.339.1155H,Dalal+2005,Hoekstra+2011,Oguri+Takada2011}
\begin{equation}
\label{eq:cl}
C^{\rm lss}_{ij}
= \int\!\frac{l\,dl}{2\pi}\,
C^{\kappa\kappa}(l)\, \hat{J}_0(l\theta_i)\hat{J}_0(l\theta_j).
\end{equation}
Here $C^{\kappa\kappa}(l)$ is the weak-lensing power spectrum as a
function of angular multipole $l$ evaluated
for a given source population and a cosmology, and
$\hat{J}_0(l\theta_i)$ is the Bessel function of the first kind and
order zero averaged over the $i$th annulus between $\theta_{i,1}$ and
$\theta_{i,2} (>\theta_{i,1})$, given as 
\begin{equation}
\hat{J}_0(l\theta_i)=\frac{2}{(l\theta_{i,2})^2-(l\theta_{i,1})^2}
\left[
l\theta_{i,2}J_1(l\theta_{i,2})-l\theta_{i,1}J_1(l\theta_{i,1})
\right].
\end{equation}
We will assume the concordance $\Lambda$CDM cosmological model of
\citet{Komatsu+2011_WMAP7} and use the fitting formula of \citet{PD96}
to compute the nonlinear mass power spectrum that enters in
equation (\ref{eq:cl}).

\subsection{Stacked Lensing Analysis}
\label{subsec:stack}

The utility of high-quality data is ultimately limited by the cosmic
 noise from large scale structure along the line of sight,  producing
 covariance between radial bins, particularly behind the cluster center,
 where magnified sources lie at greater distances. This noise is
 correlated between radial bins, but can be overcome by stacking an
 ensemble of clusters along independent lines of sight. Stacking also
 helps average over the effects of cluster asphericity and substructure
 inherent in projected lensing measurements. 
The statistical precision can be greatly improved by stacking together a
 number of clusters, especially on small angular scales
 \citep[see][]{Okabe+2010_WL}, allowing a tighter comparison of the
 averaged profile with theoretical models.

With the full mass profiles of individual clusters from combined weak
and strong lensing (Section \ref{subsec:wl+sl}), 
we can stack the clusters to 
produce an averaged radial mass profile.
Here we re-evaluate the mass profiles of the individual clusters in
$M$ logarithmically-spaced radial bins in the range $R=[R_{\rm
min},R_{\rm max}]$ 
following the prescription given in
\citet{Umetsu+2011}. 
Since the noise in different clusters is uncorrelated,
the mass profiles of individual clusters can be co-added according to 
\citep{Umetsu+2011}
\begin{equation}
\label{eq:stack}
\langle \bSigma\rangle = 
\left(\displaystyle\sum_n {\cal W}_n \right)^{-1}
 \,
\left(
\displaystyle\sum_n{ {\cal W}_n \bSigma_n}
\right), 
\end{equation}
where the index $n$ runs over all clusters, 
$\bSigma_n$ is a vector containing the discrete surface mass density
profile for the $n$th cluster, and
${\cal W}_n$ is the window matrix defined as
\begin{equation}
({\cal W}_n)_{ij} \equiv \left(C_n^{-1}\right)_{ij} (w_n)_i (w_n)_j
\end{equation}
with $(C_n)_{ij}$ and $(w_n)_i$ $(i=1,2,...,M)$ being the full
covariance matrix and the 
lensing efficiency function for the $n$th  cluster, respectively.
The error covariance matrix for the stacked mass profile
$\langle\bSigma\rangle$ is obtained as 
\begin{equation}
\label{eq:covar_stack}
{\cal C} =\left(
\displaystyle \sum_n {\cal W}_n\right)^{-1}.
\end{equation}

\section{Applications: Hubble and Subaru Observations of Four High-Mass
 Clusters} 
\label{sec:app}

\subsection{Cluster Sample and Lensing Data}
\label{subsec:data}

Following the methodology outlined in Section \ref{sec:method}, we
analyze our consistent weak and strong lensing measurements presented in
\citet{Umetsu+2011} to examine the underlying projected mass profile
$\Sigma(R)$ of a sample of four well-studied high-mass clusters
($M\simgt 10^{15}M_\odot$) at intermediate redshifts, A1689 ($z=0.183$),
A1703 ($z=0.281$), A370 ($z=0.375$), and Cl0024+17
($z=0.395$)\footnote{Careful examination of lensing, X-ray, and 
dynamical data strongly suggest that Cl0024+17 is the results of a
high-speed, line-of-sight collision of two massive clusters viewed
approximately 2--3\,Gyr after impact when the gravitational potential
has had time to relax in the center, but before the gas has recovered
\citep[see][and references therein]{Umetsu+2010_CL0024}}.  The massive
clusters we have analyzed are well-known strong lensing clusters,
displaying prominent strong-lensing features and large Einstein radii
of $\theta_{\rm Ein}\simgt 30\arcsec$ \citep[e.g., for a fiducial
source redshift $z_s\sim 2$;][]{Broadhurst+Barkana2008,Oguri+Blandford2009,Zitrin+2011_Ein}.
%
Table \ref{tab:data} gives a summary of the basic properties of the
clusters in our sample.

For these clusters, the central mass distributions ($R\simlt
200$\,kpc\,$h^{-1}$) have been recovered in detail by our strong-lensing
analysis
\citep{2005ApJ...621...53B,Zitrin+2009_CL0024,Zitrin+2010_A1703,Umetsu+2011} 
based on many sets of multiply-lensed images identified
previously in very deep multicolor imaging with {\it HST}/ACS  
\citep[e.g.,][]{2005ApJ...621...53B,Limousin+2008_A1703,Richard+2009_A1703,Richard+2010_A370,Zitrin+2009_CL0024,Zitrin+2010_A1703}.  
\citet{Umetsu+2011} developed and applied a Bayesian method to
derive model-independent projected mass profiles for five high-mass clusters
(including RXJ1347-11 in addition to the four clusters)
from Subaru weak-lensing distortion and  magnification measurements, the
combination of which can unambiguously break the mass-sheet degeneracy
inherent in any mass inversion method based solely on shape distortion data.
It was shown that for the four clusters of the present sample 
our independent strong and weak lensing mass profiles are in full agreement
in the region of overlap ($R\sim 150\,$kpc\,$h^{-1}$), and together can
be well described by, within the noise, a generalized form of the NFW
profile for CDM-dominated equilibrium halos.
This motivates us to reexamine in detail the form of the radial mass
profile for the entire clusters.

\subsection{Results}
\label{subsec:results}

 
\begin{deluxetable*}{ccccccccc} 
\tabletypesize{\scriptsize}
\tablecolumns{7}
\tablecaption{\label{tab:data} Cluster sample and lensing data} 
\tablewidth{0pt} 
\tablehead{ 
 \multicolumn{1}{c}{Cluster} &
 \multicolumn{1}{c}{Redshift} &
 \multicolumn{1}{c}{Einstein radius} &
 \multicolumn{2}{c}{Strong lensing} &
 \multicolumn{2}{c}{Weak lensing} &
 \multicolumn{1}{c}{S/N}
\\
 \multicolumn{1}{c}{} &
 \multicolumn{1}{c}{$z$} &
 \multicolumn{1}{c}{$\theta_{\rm ein}$} &
 \multicolumn{1}{c}{$R_{\rm min}^{\rm sl},R_{\rm max}^{\rm sl}$} &
 \multicolumn{1}{c}{$N^{\rm sl}$} &
 \multicolumn{1}{c}{$R_{\rm min}^{\rm wl},R_{\rm max}^{\rm wl}$} &
 \multicolumn{1}{c}{$N^{\rm wl}$} &
 \multicolumn{1}{c}{} & 
\\
 \multicolumn{1}{c}{} &
 \multicolumn{1}{c}{} &
 \multicolumn{1}{c}{($\arcsec$)} &
 \multicolumn{1}{c}{(kpc\,$h^{-1}$)} &
 \multicolumn{1}{c}{} &
 \multicolumn{1}{c}{(kpc\,$h^{-1}$)} &
 \multicolumn{1}{c}{} & 
 \multicolumn{1}{c}{}
}
\startdata 
 A1689 &
 $0.183$ &
 $53\pm 3\arcsec (z_s=3.04)$ &
 $40,125$ &
 12 &
 $129,2325$ &
 11 &
 35
\\
 A1703  & 
 $0.281$ &
 $31\pm 3\arcsec (z_s=2.627)$ &
 $40,177$   &
 14 &
 $179,2859$ &
 10 &
 29
\\
 A370 &
 $0.375$ &
 $37\pm 3\arcsec (z_s=2)$ &
 $40,149$ &
 15 &
 $152,3469$ &
 14 &
 29 
\\
 Cl0024+17  & 
 $0.395$ &
 $30\pm 3\arcsec (z_s=1.675)$ &
 $40,126$ &
 14  &
 $134,3359$ &
 12 &
 26 \\
\enddata
\tablecomments{
For each cluster a joint mass profile is defined in $N\equiv N^{\rm
 sl}+N^{\rm wl}$ discrete radial bins over the radial range of
 $R=[R_{\rm  min}^{\rm sl},R_{\rm max}^{\rm wl}]$.}
\end{deluxetable*}


\begin{figure*}[!htb] 
 \begin{center}
 \includegraphics[width=120mm,angle=270]{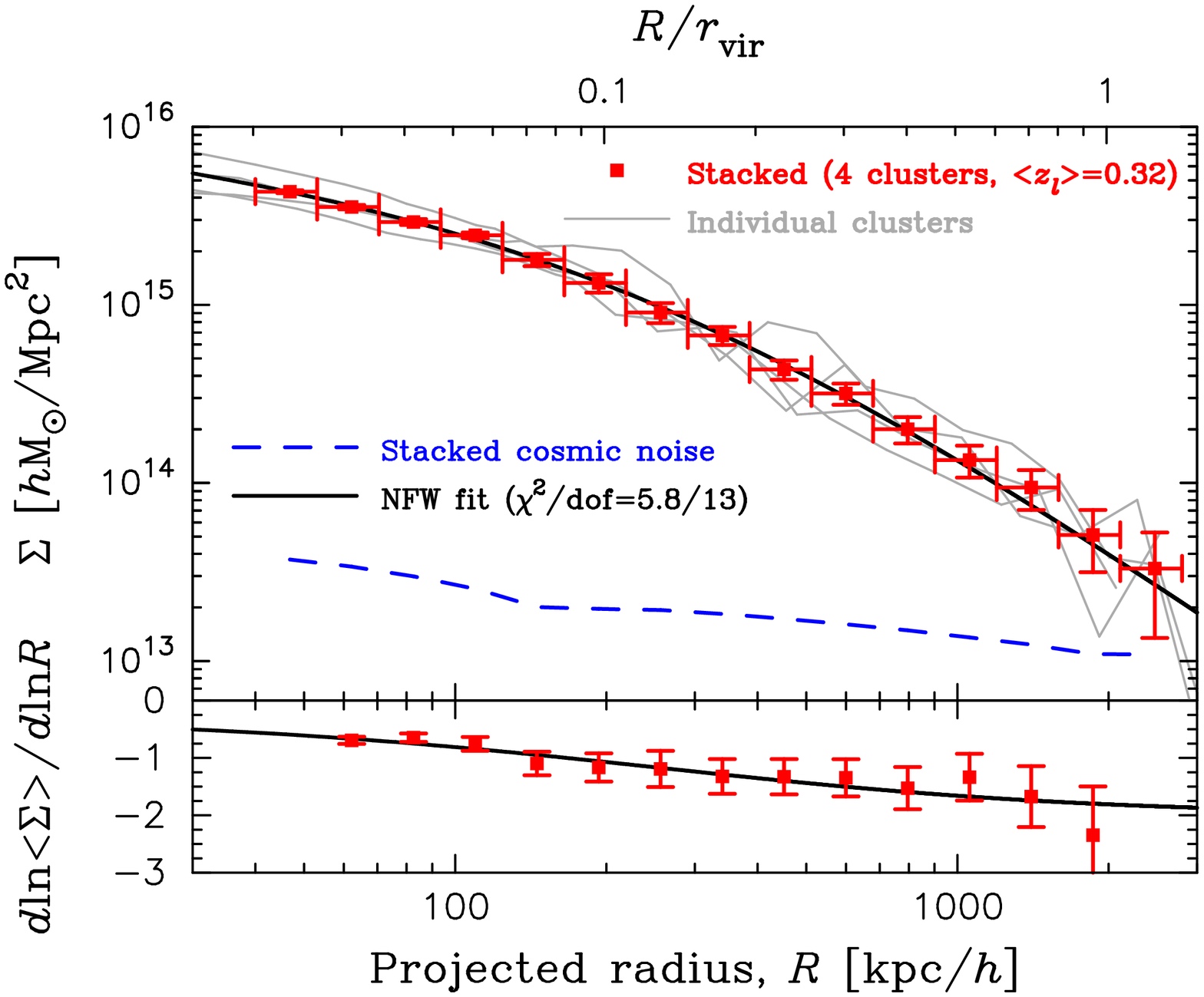} 
 \end{center}
\caption{
Top: the average projected mass profile $\Sigma(R)$ (filled squares) with its 
statistical $1\sigma$ uncertainty as a function of the projected radius
 $R$, which is obtained by stacking individual full mass profiles (thin
 gray lines) 
of four high-mass clusters
(A1689, A1703, A370, and Cl0024+17 with $M_{\rm vir}>10^{15}M_\odot$
at $\langle  z_l\rangle=0.32$)
derived from Hubble strong lensing
 ($R\simlt  150$\,kpc\,$h^{-1}$) and Subaru weak lensing ($R\simgt
 150\,$kpc\,$h^{-1}$) measurements. 
The stacked mass profile exhibits clear continuous steepening over a wide
 range of radii, from $R=40\,$kpc\,$h^{-1}$ to 
 $2800\,$kpc\,$h^{-1} \approx 1.4 r_{\rm vir}$, which is well described
 by a single NFW profile (solid line).  The dashed line shows the
 contribution to the variance from uncorrelated large scale structure
 projected along the line of sight. 
Bottom: the logarithmic slope of the stacked mass profile
(open squares with error bars), $d\ln\langle\Sigma\rangle/d\ln{R}$,
is shown as a function of projected radius along with the NFW model
 (solid line)
 shown in the top panel.  The projected logarithmic slope shows a clear
 continuous steepening with increasing radius, consistent with the NFW
 model. 
\label{fig:stack}
} 
\end{figure*}


\begin{figure*}[!htb] 
 \begin{center}
 \includegraphics[width=90mm,angle=270]{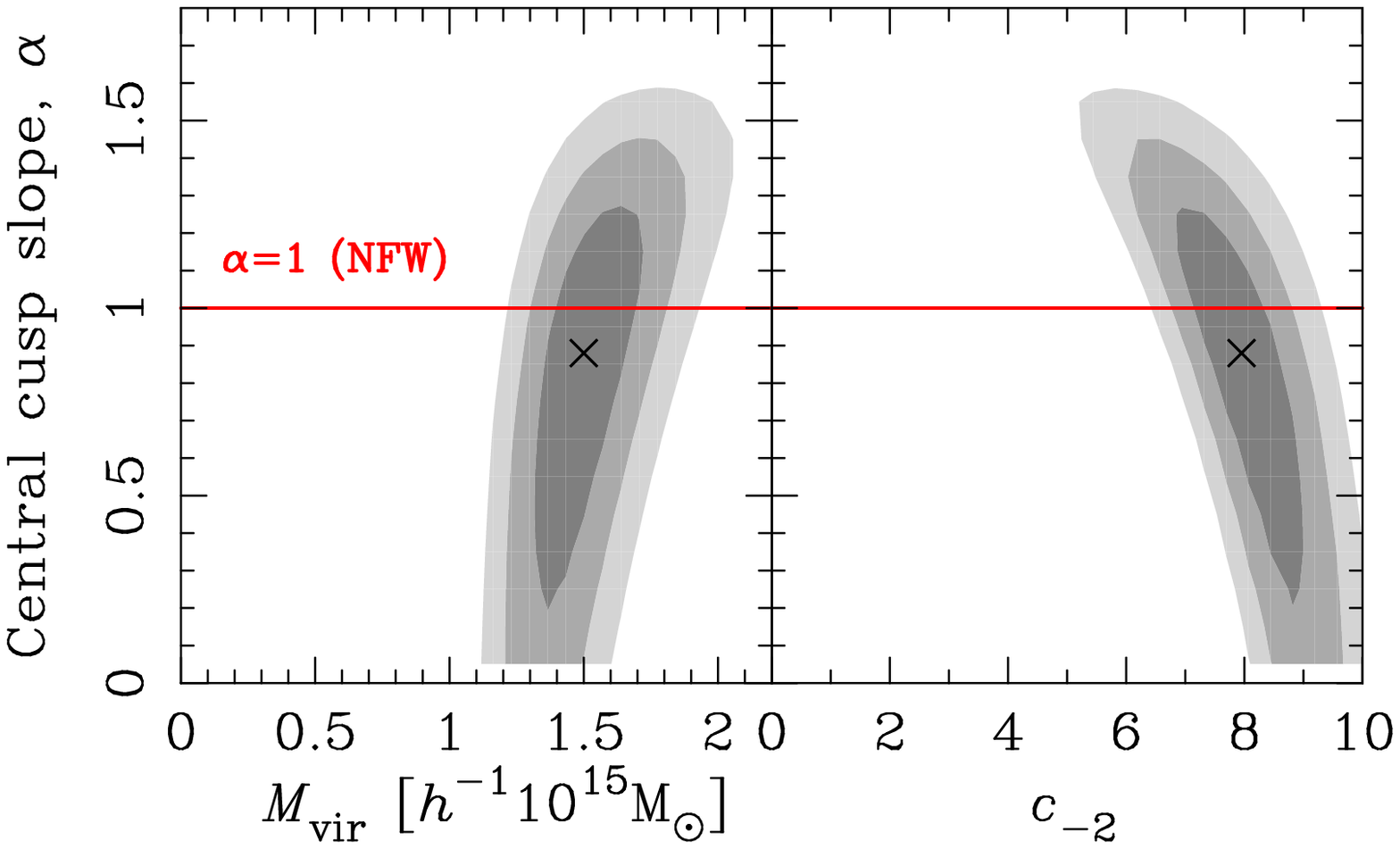} 
 \end{center}
\caption{
Constraint on the gNFW model parameters, namely, the central cusp
slope $\alpha$, the halo virial mass $M_{\rm vir}$, and the halo
concentration $c_{-2}=c_{\rm vir}/(2-\alpha)$, when all of them are
 allowed to vary, derived from the averaged radial mass profile of
 A1689, A1703, A370, and Cl0024+17 shown in Figure \ref{fig:stack}.
The left and right panels show the two-dimensional
marginalized constraints on $(M_{\rm vir},\alpha)$ and
 $(c_{-2},\alpha)$, respectively.
In each panel of the figure, the contours show the 68.3\%, 95.4\%, and
 99.7\% confidence levels, and the cross indicates the best-fit model
 parameters. 
\label{fig:2dconf}
} 
\end{figure*} 


\begin{figure*}[!htb] 
 \begin{center}
 \includegraphics[width=120mm,angle=270]{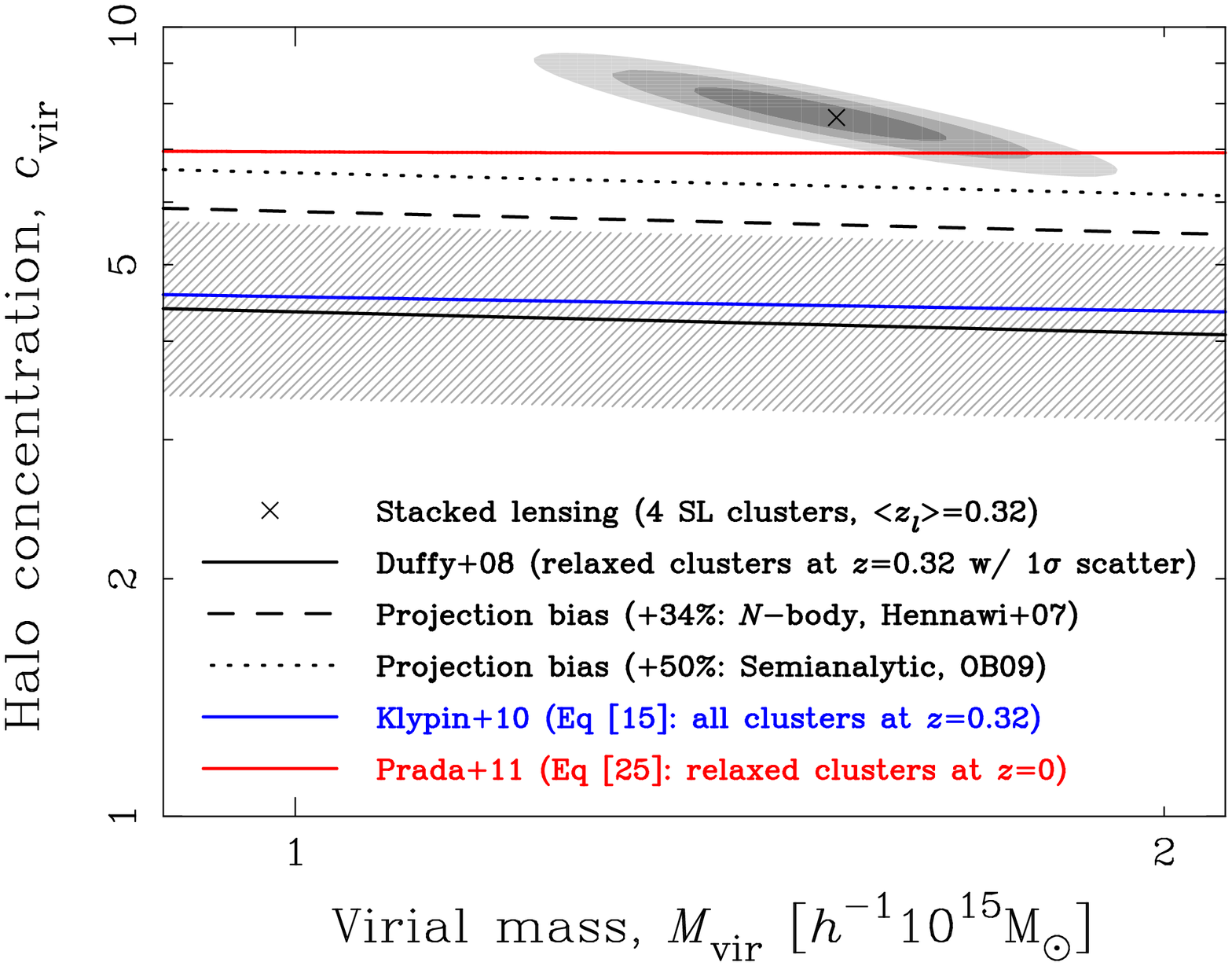} 
 \end{center}
\caption{ Joint constraints on the mass and concentration parameters
$(M_{\rm vir},c_{\rm vir})$ for a sample of four high-mass lensing
clusters (A1689, A1703, A370, and Cl0024+17) derived from their stacked
full mass profile $\langle\Sigma(R)\rangle$ (Figure \ref{fig:stack}),
compared to $\Lambda$CDM predictions
\citep{Duffy+2008,Klypin+2010,Prada+2011} in the $c_{\rm vir}$--$M_{\rm
vir}$ plane.  The cross shows the best-fit NFW parameters, and the
contours show the 68\%, 95\%, and 99.7\% confidence levels
($\Delta\chi^2=2.3, 6.17$, and $11.8$).  The $N$-body predictions of
\citet[][]{Duffy+2008}, \citet[][]{Klypin+2010}, and \citet{Prada+2011}
are shown as solid curves, with $1\sigma$ lognormal scatter \citep[taken
from][]{Duffy+2008} indicated by the shaded area.  Also shown are the
levels of selection and projection bias for a strong-lensing cluster
population derived from $N$-body ($34\%$; dotted line) and
semi-analytical ($50\%$; dashed line) simulations, where the prediction
by \citet{Duffy+2008} is taken as the reference of the comparison.
\label{fig:cmplot} }
\end{figure*}

Our weak and strong lensing data together cover a wide range of radius
ranging typically from $R\sim 10\,$kpc$\,h^{-1}$ to
$2000$--$3500$\,kpc$\,h^{-1}$ \citep{Umetsu+2011}, depending on the
cluster redshift as limited by the field of view of Subaru/Suprime-Cam
($34\arcmin \times 27\arcmin$). 
Table \ref{tab:data} lists for each cluster the radial ranges
$R=[R_{\rm  min}^{\rm sl},R_{\rm max}^{\rm sl}]$
and
$[R_{\rm  min}^{\rm wl},R_{\rm max}^{\rm wl}]$
of strong and weak lensing measurements, respectively, used to define a
joint discrete mass profile $\bSigma =\{\Sigma(R_i)\}_{i=1}^{N}$, given
in a total of $N$ radial bins spanning from $R_{\rm
min}=R_{\rm min}^{\rm sl}$ to $R_{\rm max}=R_{\rm max}^{\rm wl}$.
In Table \ref{tab:data}, we also quote values of the total ${\rm S/N}$
ratio in our joint cluster mass profiles ($\bSigma$) 
obtained using the full covariance matrix $C$.
We find that ignoring the cosmic noise contribution (equation
[\ref{eq:cl}]) will underestimate the errors by $\sim 30\%$--$40\%$.
To evaluate $C^{\rm lss}$ for strong-lensing observations, 
we projected the matter power spectrum out to a fiducial depth of
$z_s=2$, which is a typical source redshift  
of strongly-lensed arcs in clusters at intermediate redshifts.
We used the estimated mean source redshifts given in Table 3 of
\citet{Umetsu+2011} for weak lensing.

We show in the top panel of Figure \ref{fig:stack} the resulting
averaged radial mass profile $\langle \Sigma(R)\rangle$ 
in $M=15$ logarithmically-spaced bins
with its
statistical $1\sigma$ uncertainty (given as the square root of the
diagonal part of the full covariance matrix ${\cal C}$), 
obtained by stacking the four clusters using
equations (\ref{eq:stack}) and (\ref{eq:covar_stack}).
Note,
no scaling has been applied to match the mass normalizations
between the four clusters, which span a relatively narrow range in mass,  
$1.3\simlt M_{\rm vir}/(10^{15} M_\odot\,h^{-1})\simlt 2.3$ 
\citep[see Table 6 of][]{Umetsu+2011}.
For our sample, we find a sensitivity-weighted average cluster redshift 
of $\langle z_l\rangle \simeq 0.32$, which is fairly close to
the simple average of $\overline{z_l}=0.31$ due to the narrow
redshift coverage of our cluster sample.
The stacked mass profile exhibits a smooth radial trend with a
clear radial curvature
over a wide range of radius from $R=40$\,kpc\,$h^{-1}$ to
$2800$\,kpc\,$h^{-1} \approx 1.4 r_{\rm vir}$,
and is detected at a high significance level of $58\sigma$, with the
contribution from cosmic covariance included.
Here the maximum radius for the stacking analysis represents approximately
the average maximum radius $\langle R_{\rm max}\rangle$ covered by our
data. 
Also shown in Figure \ref{fig:stack} is the cosmic noise contribution,
which increases toward the cluster center.
A noticeable increase of the stacked cosmic noise is seen at $R\simlt
150\,$kpc\,$h^{-1}$, within which the averaged profile is dominated by
strong lensing measurements with greater depth.
In the bottom panel of Figure
\ref{fig:stack}, we plot the logarithmic density slope
$\gamma_{\rm 2D}(R)\equiv -d\ln{\langle\Sigma\rangle}/d\ln{R}$ of the
stacked mass profile.
The logarithmic gradient of the average profile shows a clear 
continuous steepening with increasing radius in projection.

To quantify and characterize the averaged cluster mass distribution,
we compare the $\langle\Sigma\rangle$ profile with 
the physically and observationally motivated NFW model.
Here we consider a generalized parametrization of the
NFW model (gNFW, hereafter) of the form 
\citep{Zhao1996,Jing+Suto2000}:
\begin{equation}
\label{eq:gnfw}
\rho(r)=\frac{\rho_s}{(r/r_s)^\alpha(1+r/r_s)^{3-\alpha}},
\end{equation} 
where $\rho_s$ is the characteristic density, 
$r_s$ is the characteristic scale radius, and
$\alpha$ represents the inner slope of the density profile.
This reduces to the NFW model for $\alpha=1$.
We introduce the radius $r_{-2}$ at which the logarithmic
slope of the density is isothermal, i.e., $\gamma_{\rm 3D}=2$. For the
gNFW profile, $r_{-2}=(2-\alpha)r_s$, and thus the corresponding
concentration parameter reduces to $c_{-2}\equiv r_{\rm
vir}/r_{-2}=c_{\rm vir}/(2-\alpha)$.
We specify the gNFW model with the central cusp slope, $\alpha$, 
the halo virial mass, $M_{\rm vir}$, and the concentration,
$c_{-2}=c_{\rm vir}/(2-\alpha)$.
We employ the radial dependence of the gNFW lensing profiles given by
\citet{Keeton2001_mass}.

First, when the central cusp slope is fixed to $\alpha=1$ (NFW), the
best-fit model for the averaged $\langle\Sigma\rangle$ profile is
obtained as $M_{\rm vir}=1.54^{+0.11}_{-0.10}\times
10^{15}M_\odot\,h^{-1}$ and $c_{-2}=c_{\rm vir}=7.68^{+0.42}_{-0.40}$
with the minimized $\chi^2$ value ($\chi^2_{\rm min}$) of 5.8 for 13
degrees of freedom (dof), corresponding to a $Q$-value goodness-of-fit
of $Q=0.952$. 
This model yields an Einstein radius of $\theta_{\rm
Ein}=39.9\arcsec^{+4.4}_{-4.1}$ for a fiducial source at $z_s=3$.   
The resulting best-fit NFW parameters from the stacked analysis
are consistent with the respective sample weighted means of
the individual NFW model fits obtained by \citet[][Table
6]{Umetsu+2011}: $\langle M_{\rm vir}\rangle =1.44\pm 0.11 
\times 10^{15}M_\odot\,h^{-1}$ and $\langle c_{\rm vir}\rangle = 7.76\pm
0.79$.
Next, when $\alpha$ is allowed to vary, a gNFW fit to
$\langle\bSigma\rangle$ gives $M_{\rm vir}=1.50^{+0.14}_{-0.13}\times
10^{15}M_\odot\,h^{-1}$, $c_{-2}=7.91^{+0.72}_{-0.75}$, and
$\alpha=0.89^{+0.27}_{-0.39}$ with $\chi^2_{\rm min}/{\rm dof}=5.7/12$
and $Q=0.931$ ($\theta_{\rm Ein}=38.4\arcsec^{+12.2}_{-10.2}$ at
$z_s=3$), being consistent with a simple NFW model with $\alpha=1$.
Thus the addition of the $\alpha$ parameter does not improve the fit
substantially, as shown by the quoted $\chi^2$ and $Q$ values \citep[see
also][]{Zitrin+2011_A383}.  The two-dimensional marginalized constraints
($68.3\%, 95.4\%$, and $99.7\%$ confidence levels) on $(M_{\rm
vir},\alpha)$ and $(c_{-2},\alpha)$ are shown in Figure
\ref{fig:2dconf}.
Finally, a force fit to the singular
isothermal sphere (SIS) model ($\rho\propto r^{-2}$) yields a poor fit
with $\chi^2_{\rm min}/{\rm dof}=78.5/14$, so that the SIS model is strongly
disfavored at $62\sigma$ significance from a likelihood-ratio test,
based on the difference between $\chi^2$ values of the best-fit NFW and
SIS models: $\Delta\chi^2\equiv \chi^2_{\rm SIS,min}-\chi^2_{\rm
NFW,min}=72.6$ for a 1 degree-of-freedom.

\section{Discussion and Conclusions}
\label{sec:discussion}

We have developed a method for improving the statistical precision of
cluster mass profiles, combining independent weak-lensing distortion,
magnification, and strong-lensing measurements.  This extends recent
weak-lensing work by \citet{Umetsu+2011} to include the central
strong-lensing information in a stacking analysis, for full radial
coverage.  Our methods take into account the cosmic covariance from
uncorrelated large scale structure projected along the line of sight
\citep{2003MNRAS.339.1155H,Hoekstra+2011}, as well as the effect of
different cluster redshifts, so that error propagation in terms of 
lensing efficiency of individual clusters can be properly averaged.

We have applied our method to a sample of four similarly high-mass lensing
clusters (A1689, A1703, A370, and Cl0024+17), for which we have
previously identified multiply-lensed images  and measured weak
magnification and distortion effects from deep {\it HST} and Subaru
observations
\citep{2005ApJ...621...53B,UB2008,Umetsu+2010_CL0024,Zitrin+2010_A1703,Medezinski+2010,Medezinski+2011,Umetsu+2011}. 
For our sample of massive clusters the strong and weak lensing regimes
contribute equal logarithmic coverage of the radial profile and can be
compared for consistency in the region of overlap. 
We have formed an averaged radial mass profile $\langle\Sigma(R) \rangle$ 
from stacking the clusters (Figure \ref{fig:stack}),
which shows a progressive steepening  with increasing radius from
$R=40\,$kpc\,$h^{-1}$ to $2800\,$kpc\,$h^{-1}$.
The inner radial boundary is chosen to be sufficiently large to avoid
smoothing from  cluster miscentering effects
\citep{Johnston+2007_SDSS1}, where the typical offset between the BCG
and the dark matter center is estimated as $d\simlt 20\,$kpc\,$h^{-1}$
for our sample from our detailed strong-lens modeling (see Section
\ref{subsec:wl+sl}).
The stacked full mass profile is detected at a high significance level
of $58\sigma$ over the entire radial range.
It is found here that ignoring the cosmic noise contribution will
underestimate the errors by $\sim 30\%$--$40\%$. This is due to the 
correlation of this noise between radial bins and can only be reduced by 
averaging over independent lines of sight, with uncorrelated line of
sight structures, i.e. by averaging over well separated clusters.  

Our stacked projected mass profile with a continuously steepening radial
trend is very accurately described by the NFW form predicted for the
family of CDM-dominated halos, whereas it strongly disfavors the SIS
model at $62\sigma$ significance.
In the context of an assumed gNFW profile, the central cusp slope is
constrained as $\alpha=0.89^{+0.27}_{-0.39}$ (at $r\simgt 0.02 r_{\rm
vir}$; see Figures \ref{fig:stack} and \ref{fig:2dconf}), being
consistent with, but slightly shallower than, the simple NFW form with
$\alpha=1$.  Our results are in agreement with recent high-resolution
simulations, which find asymptotic inner slopes somewhat shallower than
unity, $\gamma_{\rm 3D}(r\to 0)\simlt 0.9$, for galaxy- and
cluster-sized $\Lambda$CDM halos 
\citep[e.g.,][]{Merritt+2006,Graham+2006,Navarro+2010}.  
Note NFW define this profile for halos which they identify as in virial
equilibrium, in terms of the simulated CDM particles
\cite[see Section 2.2.2 of][]{1997ApJ...490..493N}.
The clusters we have selected for our stacked analysis are, in terms of
their lensing properties, very well behaved with at most only $\sim 10\%$
perturbations in mass visible locally in the two-dimensional mass
distribution, and otherwise very symmetric over most of the radius
\citep{2005ApJ...621...53B,BTU+05,BUM+08,Umetsu+2010_CL0024}. 
Detailed hydrodynamical simulations show
that equilibrium is relatively rapidly achieved in only a few sound
crossing times after a major merger, though some dynamical and gas
disruption may continue for over a Gyr. This is not important in terms
of the central relaxation time of the dark matter
\citep{2001ApJ...561..621R,Umetsu+2010_CL0024}. 

An accurate measurement of the cluster mass profile enables us
to constrain dark matter models.
Recently \citet{BEC2009} examined in detail an extremely light bosonic
dark matter (ELBDM) model ($m\sim 10^{22}$\,eV) as an alternative to
CDM in the context of nonlinear cosmic structure formation.  ELBDM
with a de-Broglie wavelength of astronomical length scales, if it
exists,
may well be in a ground-state Bose-Einstein
condensate and hence well described by a coherent wave function,
which may naturally account for the perceived lack of small galaxies
relative to the $\Lambda$CDM model \citep{Klypin+1999,Peebles+Nusser2010}.
\citet{BEC2009} showed that, irrespective of whether halos form through
accretion or merger,  
ELBDM halos can form steepening density profiles 
of the form similar to the standard CDM,  but with perhaps a
steeper central cusp slope of $\gamma_{\rm 3D}\simeq 1.4$ and a
shallower outer slope of $\gamma_{\rm 3D}\simeq 2.5$. 
During a merger
between condensates interesting large-scale interference occurs which
will differ markedly from standard collisionless CDM, and it will be
important to explore this class of dark matter further via more
extensive and detailed simulations for testing against accurate lensing
profiles of both relaxed and merging clusters.

The mean concentration for the four massive lensing clusters
considered here is found to be $c_{\rm vir}=7.68^{+0.42}_{-0.40}$ (at
a mean virial mass $M_{\rm vir}=1.54^{+0.11}_{-0.10}\times
10^{15}M_\odot\,h^{-1}$), which is apparently higher than the standard
$\Lambda$CDM predictions evaluated at the mean redshift $\langle
z_l\rangle=0.32$ of our sample: $c_{\rm vir}=4.5^{+1.3}_{-1.0}$ (the
errors quoted represent a $1\sigma$ lognormal scatter of
$\sigma[\log_{10}{c_{\rm vir}}]=0.11$) for relaxed clusters derived by
\citet{Duffy+2008} from $N$-body simulations based on the WMAP 5-year
data and $c_{\rm vir}\approx 4.4$ by \citet{Klypin+2010} from the
recent {\it Bolshoi} $\Lambda$CDM $N$-body simulation.  
More recent results with greater mass resolution based on four large
$N$-body simulations  (Bolshoi, MultiDark, Millennium-I and II) exhibit
a complex mass and redshift dependence of the median concentration,
namely a flattening and upturn of concentration at very high mass and redshift
\citep{Prada+2011}.  Accordingly, their concentrations derived for
cluster-sized halos (i.e., rare objects corresponding to high-$\sigma$
peaks in the primordial density field) are substantially higher than 
previous results based on smaller simulations.   Interestingly, they
find a concentration of $c_{\rm vir}\sim 7$ for their most-massive {\it
relaxed} halos with $M_{\rm vir}\approx 10^{15}M_\odot \,h^{-1}$ at
$z=0$ \citep[Figure 15 of][]{Prada+2011}. 
A comparison between our results and the $\Lambda$CDM predictions
\citep{Duffy+2008,Klypin+2010,Prada+2011} is given in
Figure \ref{fig:cmplot}.

An accurate characterization of the observed sample is crucial for any
cluster-based cosmological tests. 
In the extreme case,
those clusters identified by the presence of a giant arc represent the
most lensing-biased population.
Calculations of the enhancement of the
projected mass and hence boosted Einstein radii (say, $\theta_{\rm
Ein}>20\arcsec$) find a statistical bias of $\sim 34\%$
derived from $N$-body simulations of the $\Lambda$CDM model
\citep{2007ApJ...654..714H}.  Semi-analytical simulations incorporating
idealized triaxial halos yield a $\sim 50\%$ bias correction
\citep{Oguri+Blandford2009}. 
Applying a conservative $50\%$ bias correction,
we find a discrepancy of about $1.8\sigma$ with respect to the
$\Lambda$CDM predictions by the \citet{Duffy+2008} model
for relaxed clusters (see Figure
\ref{fig:cmplot}). If this 
large bias ($\sim 50\%$) is coupled to a sizable intrinsic scatter in
concentration, estimated for the full halo population to be
$\sigma[\log_{10}c_{\rm   vir}]=0.11$--$0.15$, then our measurements can
come into line with standard $\Lambda$CDM.

The results presented here are very favorable in terms of the standard
explanation for dark matter, as collisionless and non-relativistic,
interacting only via gravity, with a very precise match between our
composite mean mass profile, and that of the general form of the mass
profile advocated for massive halos in virial equilibrium.  The
relatively high concentration we obtain for the averaged profile is
consistent with previous lensing work which similarly detected a
concentration excess in the lensing based measurements for many
individual relaxed strong-lensing clusters
\citep[e.g.,][]{2003A&A...403...11G,2003ApJ...598..804K,2007MNRAS.379..190C,BUM+08,Oguri+2009_Subaru}.
This possibly interesting tension between cluster lensing observations
and $\Lambda$CDM models can be more definitively addressed with
full-lensing data for new cluster surveys, such as
CLASH\footnote{Cluster Lensing And Supernova Survey with Hubble (P.I.:
M. Postman),
http://www.stsci.edu/{\large\textasciitilde{}}postman/CLASH/}, LoCuSS,
Subaru Hyper Suprime-Cam, and XMM-XXL \citep{XXL2010}, to meaningfully
examine the $c_{\rm vir}$--$M_{\rm vir}$ relation over a wider mass and
redshift range when applied to sizable samples of relaxed clusters. It
is highly desirable to cover the full profile by combining accurate weak
and strong lensing measurements, requiring several sets of multiple
images over a wide range of source redshift, to obtain a meaningful
model-independent inner profile and to add weak lensing with sufficient
color information to exclude the otherwise sizable dilution effect on
the weak lensing signal from foreground and cluster members.  The CLASH
survey is in particular designed to generate such useful data free of
systematics in both the weak and strong regime, with first results for
the substantial smaller mass cluster A383 with 
$M_{\rm vir}= 5.37^{+0.70}_{-0.63} \times 10^{14}M_\odot\,h^{-1}$
\citep{Zitrin+2011_A383} showing similar behavior ($c_{\rm
vir}=8.77^{+0.44}_{-0.42}$).


\acknowledgments
We thank the anonymous referee for a careful reading of the manuscript 
and and for providing useful comments. 
We are very grateful for discussions with Nobuhiro Okabe, Sandor Molnar,
Jack Sayers, and Alister Graham,
whose  comments were very helpful.  
We thank Nick Kaiser for making the IMCAT package publicly available.
The work is partially supported by the National Science Council of Taiwan
under the grant NSC97-2112-M-001-020-MY3. 
KU acknowledges support from the Academia Sinica Career Development Award.

\end{document}